\documentclass[twocolumn,aps,prl,showpacs,amsmath,amssymb,psfrac,superscriptaddress]{revtex4}
\usepackage{amsmath}

\usepackage{epsfig}
\usepackage{graphicx}
\usepackage{dcolumn}
\usepackage{bm}

\renewcommand{\dag}{^{\dagger}} 
\newcommand{\dl}{\partial_{\ell}}

\newcommand{\e}{\epsilon}
\renewcommand{\k}{{k,\sigma}}
\newcommand{\up}{\uparrow}
\newcommand{\down}{\downarrow}

\def\gl{\lower.35em\hbox{$\stackrel{\textstyle>}{\textstyle<}$}}
\def\gapp{\lower.35em\hbox{$\stackrel{\textstyle>}{\sim}$}}
\def\lapp{\lower.35em\hbox{$\stackrel{\textstyle<}{\sim}$}}

\begin{document}

\title{First Order Superfluid to Bose Metal Transition in 
Systems with Resonant Pairing}
\author{T. Stauber}
\affiliation{Instituto de Ciencia de Materiales de Madrid, CSIC, Cantoblanco, E-28049 Madrid, Spain}
\author{J. Ranninger}
\affiliation{Institut NEEL, CNRS \& Unversit\'e Joseph Fourier, BP 166, 38042 Grenoble-C\'edex 9, France}
\date{\today}
\begin{abstract}
Systems showing resonant superfluidity, driven by an exchange coupling
of strength $g$ between uncorrelated pairs of itinerant fermions and
tightly bound ones, undergo a first order phase transition as $g$
increases beyond some critical value $g_c$. The superfluid phase for
$g \leq g_c$ is characterized by a gap in the fermionic single
particle spectrum and an acoustic sound-wave like collective mode of
the bosonic resonating fermion pairs inside this gap.  For $g>g_c$
this state gives way to a phase uncorrelated bosonic liquid with a
$q^2$ spectrum.
\end{abstract}
%
%
\maketitle
{\it Introduction.} The issue of the crossover between a BCS type superfluid 
and a Bose Einstein condensation (BEC) of tightly bound fermion pairs has 
received increasing attention ever since  it became 
feasible to experimentally control such a crossover in fermionic 
atomic gases in optical traps and lattices\cite{atomgasBEC}. From the 
theoretical side, 
such a crossover had until then been mainly studied for single 
component fermionic systems with  static, short-range 
attractive inter-particle 
interaction\cite{BCS-BEC}. It became clear that in these systems one is 
confronted with a unitarity limit problem, where the scattering length 
abruptly changes 
from positive to negative infinity. Numerical as well as  analytical 
results\cite{unitarity} suggest that around such a  unitarity 
point, one should expect, effectively, two kinds of quasi-particles 
coexisting with each other: itinerant fermions and 
tightly bound fermion pairs behaving as bosons. 

Such a situation had been anticipated in connexion with the physics in
electron-lattice coupled systems in a regime between weak and strong
coupling and resulted in the proposition of the boson-fermion model
(BFM)\cite{Ranninger-Robaszkiewicz-85}. The essence of the BFM
scenario is to assume itinerant fermions and localized bosons (made up
of two tightly bound fermions) - the two species which represent the
physics of the extreme weak and strong coupling limits of such systems. The
crossover behavior between those two limiting regimes is then
parametrized by an exchange coupling of strength $g$ between the
two species. This scenario can be
justified for: (i) fermionic atomic gases, where pairs of fermions exist in
different electron-nuclear spin configurations, favoring or disfavoring binding
between them (magnetic field tunable Feshbach
resonance)\cite{Feshbach}, (ii) hole pairing in strongly correlated
systems, showing plaquette RVB states\cite{Altman-Auerbach-02} and
(iii) resonating bipolarons in moderately strongly coupled
electron-lattice systems\cite{Ranninger-Romano-06}, for which this
scenario was anticipated in the first
place\cite{Ranninger-Robaszkiewicz-85}.

Contrary to fermionic systems with attractive inter-particle
interaction, where the ground state is always superfluid, the BFM
scenario shows a superfluid (amplitude driven) BCS-like state for weak
coupling, while for strong coupling it represents a liquid of
spatially phase uncorrelated bonding pairs:
$1/\sqrt{2}(u_ic_{i\uparrow}^{\dagger}c_{i\downarrow}^{\dagger} +
v_ib_i^{\dagger})$.  $u_i= |u_i|e^{i\phi_i/2}, v_i =
|v_i|e^{-i\phi_i/2}$ (with $|u_i|^2 +|v_i|^2=1$) denote the local
amplitudes and phases of the two components, and
$c_{i\sigma}^{\dagger}$, respectively $b_i^{\dagger}$, the creation
operators of the itinerant fermions (with spin $\sigma$) and of the localized
bosonic tightly bound fermion pairs at site $i$. The emergence of a
macroscopic superfluid state requires the onset of
long-range phaselocking of such bonding pairs. This needs a
concomitant weakening of the local phase
correlation of the bonding pairs\cite{Cuoco-Ranninger-04-06}. 
It is this competition
between local and global phaselocking which is at the origin of a
phase transition between a superfluid state for $g \leq g_c$ and a
state of spatially phase uncorrelated bonding pairs for $g > g_c$. In this Letter, we argue that this transition is of first order.

The exact nature of that phase uncorrelated state is presently not yet
fully understood. For very large values of $g$, it eventually must be
an insulator. Here, we show that close to $g_c$ it
constitutes a Bose Metal made out of such bonding pairs and present
the first study of the excitation spectra expected for such a BFM
scenario, which treats the fermions and bosons on equal footing as we
vary the exchange coupling $g$. This will be done on the basis of a
renormalization procedure for Hamiltonians\cite{Gla94,Flowequ} designed in
such a way as to optimally provide a fixed point Hamiltonian with
single-particle spectra for effective fermionic and bosonic
quasi-particles and reducing the interactions between them to higher
order correction terms such that, in a first step, they can be
neglected.

{\em The model.} The Hamiltonian for the BFM is divided into a free and a 
boson-fermion  exchange coupling interacting part, i.e., $H=H_0+H_{\rm int}$:
\begin{align}
    H_0&=\sum_\k(\e_k^\sigma -\mu)c_\k\dag c_\k+\sum_q (E_q-2\mu) b_q\dag b_q \notag \\
    H_{\rm int}&=\frac{1}{\sqrt{N}}\sum_{k,p}(g_{k,p}b_{k+p}\dag c_{k,\downarrow}c_{p,\uparrow}+g_{k,p}^*b_{k+p}c_{k,\uparrow}\dag c_{p,\downarrow}\dag)
\end{align}

Important for this model is to remember that the bosons are composed of 
two fermions. 
This is taken care of by requiring that both  types of particles have the 
same chemical potential $\mu$, fixed such that the total 
particle density $n_{\rm tot}=n_{F\uparrow}+n_{F\downarrow}+2n_B$ is 
constant. We shall in the present study consider the case of low overall 
density and choose for that purpose $n_{F,\sigma} \simeq n_B \simeq 0.25$. 

\begin{figure}[t]
  \begin{center}
    \includegraphics*[width=3in,angle=0]{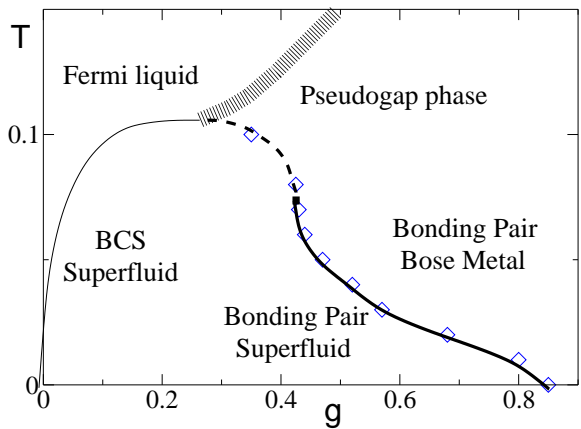}
    \caption{(Color online) Phase diagram of the BFM. The thin solid line marks the transition from a Fermi liquid to a BCS-like superfluid phase which merges into a crossover (shaded) with the pseudogap phase. The thick solid (dashed) line mark the first (second) order transition from a superfluid of locally phase correlated boson-fermion bonding pairs to an itinerant bosonic  metal (Bonding Pair Bose Metal). Diamonds represent the results of the present calculation.} 
\label{fig1}
\end{center}
\end{figure}

Fig. \ref{fig1} shows the generic phase diagram of the BFM at low densities. For small values of exchange couplings $g$, the thin solid line indicates the transition from a Fermi liquid (FL) to a BCS type superfluid phase, controlled by amplitude fluctuations and characterized by the opening of a  gap in the single particle fermion  spectrum. This transition line merges, with increasing $g$,  into the pseudogap crossover (shaded) which separates a canonical FL from a fermionic system with strong pair correlations (pseudogap phase). This crossover was analyzed in detail in Refs. \onlinecite{DR01,DR0304}. In the present work, we focus on the large exchange coupling regime and the transition from a superfluid to a Bose Metal phase of resonating fermion pairs. 

{\it The method.} The flow equation approach consists in performing continuous unitary transformations upon the initial Hamiltonian until the Hamiltonian is (block) diagonal. This is done in a systematic way by first decoupling states with high energy differences and, in the end, decoupling almost degenerate states. Previous applications of continuous unitary transformations considered the weak coupling regime of the BFM, where the zero momentum Cooper pair channel is the dominant interaction and the bosonic dispersion can be approximated by its $q=0$-value\cite{DR01}. In order to describe the evolution of the Hamiltonian in the strong-coupling phase with and without global phase coherence, it is crucial to treat the bosonic dispersion in the whole Brillouin zone.

Flow equations are formulated in terms of an anti-Hermitean generator $\eta(\ell)$ and read $\partial_\ell H(\ell)=[\eta(\ell),H(\ell)]$, with $\ell$ being the flow parameter. In this work, we shall choose the generator canonically, i. e., $\eta(\ell)=[H_0(\ell),H(\ell)]$, implying  $\dl \text{Tr}(H(\ell)-H_0(\ell))^2\leq0$ and hence $g_{k,p}\to0$ for $\ell\to\infty$ \cite{Flowequ}. The infinitesimal transformations induce a flow on the fermion as well as boson dispersion and exchange coupling constants $g_{k,p}$:
\begin{align}
\dl g_{k,p}&=-\alpha_{k,p}^2g_{k,p}\label{flowequations},\qquad \alpha_{k,p}=\e_k^\down+\e_p^\up-E_{k+p}\notag\\
\dl \e_k^\sigma&=\frac{2}{N}\sum_p\left(\alpha_{p,k}|g_{p,k}|^2\delta_{\sigma,\uparrow}+\alpha_{k,p}|g_{k,p}|^2\delta_{\sigma,\downarrow}\right)n_{k+p}^{(BE)}\notag\\
\dl E_q&=\frac{2}{N}\sum_p\{\alpha_{q-p,p}|g_{q-p,p}|^2[-1+n_{\down,q-p}^{(FD)}]\notag\\
&+\alpha_{p,q-p}|g_{p,q-p}|^2n_{\up,q-p}^{(FD)}\}
\end{align}       
 For $\ell=\infty$, the renormalized fermions and bosons are decoupled and characterized by the fixed point dispersions $\e_k^*=\e_k(\ell=\infty)$ and $E_q^*=E_q(\ell=\infty)$.

In order to close the infinite hierarchy of newly generated interaction terms, higher order interaction terms in their {\em normal ordered form} with respect to the fixed point Hamiltonian $H_0$ are neglected. To lowest order, this introduces bilinear expectation values of the fermionic and bosonic operators. We will choose the expectation values explicitly $\ell$-dependent, i. e., 
\begin{align}
&n_{k\sigma}^{(FD)}(\ell)\equiv\langle c_\k\dag c_\k\rangle(\ell)=\left(e^{(\e_k^\sigma(\ell)-\mu(\ell))/k_BT}+1\right)^{-1},\notag\\
&n_{q}^{(BE)}(\ell)\equiv\langle b_q\dag b_q\rangle(\ell)=\left(e^{(E_q(\ell)-2\mu(\ell))/k_BT}-1\right)^{-1}.\label{SCmu}
\end{align}
The chemical potential $\mu(\ell)$ is determined such that the total number of 
particles $N_{\rm tot}=\sum_{k\sigma}n_{k\sigma}^{(FD)}(\ell)+\sum_{q}n_{q}^{(BE)}(\ell)$ is conserved, which induces a flow of $\mu(\ell)$  with a fixed point value $\mu^*=\mu(\ell=\infty)$.

The justification  for assuming fermion and boson distribution functions, 
Eq. (\ref{SCmu}), in form of the distribution functions for 
non-interacting quasi-particles is the following: For small $\ell$, 
states with high energy differences are decoupled, which corresponds to the
perturbative regime and thus to well defined  fermionic 
and bosonic quasi-particles. For large $\ell$, the  exchange interaction 
$g_{k,p}$ has basically dropped to zero,  according to how this renormalization procedure was constructed. It is this asymptotic regime 
$\ell\to\infty$ which determines the critical behavior of the system 
\cite{Asymptotics}. Since the main effect of the exchange coupling $g$ is 
not to induce a significant change in the relative number of fermions or 
bosons but to introduce phase coherence between the Cooper pairs and bosons, 
Eq. (\ref{SCmu}) represents a good approximation also in the strong 
coupling regime \cite{Comment}. 
Hence, even though the truncation scheme is perturbative in the exchange coupling, the strong coupling regime can be discussed within the flow equation approach (as has been shown also for the Hubbard model, see e.g. Ref. \onlinecite{StrongC}). 
   
{\it Numerical results.} As initial conditions for the renormalization flow, we choose a dispersionless bosonic band with $E_q=\Delta_B=-0.6$ and a fermionic tight-binding dispersion $\e_\k=\e_k=-2t\cos(ka)$ where we set the lattice constant $a=1$ and use the bandwidth $D=4t$ 
as energy unit. We further assume a local exchange interaction $g_{k,p}=g$. For the numerical integration of the flow equations, Eq. (\ref{flowequations}), we choose a one-dimensional system with $N=100$ lattice sites which already resembles the thermodynamic limit.\cite{footnote} As we shall discuss below, the phase transition is determined by the  chemical potential moving out of the renormalized fermionic band of the fixed point Hamiltonian. Since the 
divergencies of the density-of-states at the bare band edges is smeared out in the course of the renormalization flow, the basic results are expected to remain qualitatively valid also in higher dimensions.

\begin{figure}[t]
  \begin{center}
    \includegraphics*[width=3in,angle=0]{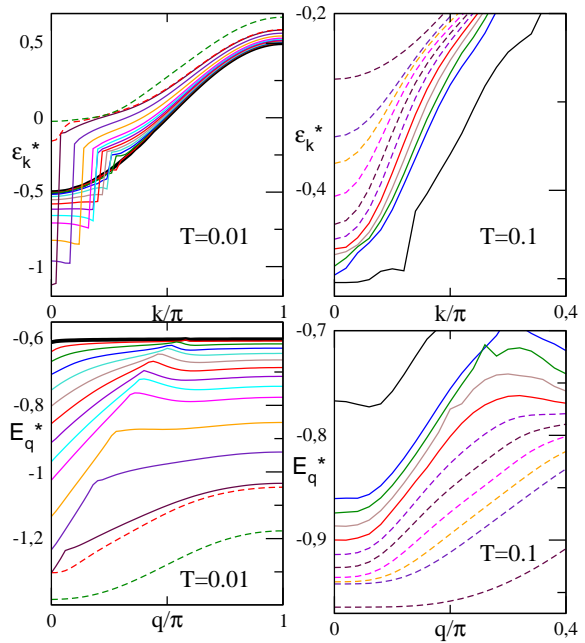}
    \caption{(Color online) The fermionic and bosonic fixed point dispersion $\epsilon_q^*$ (upper panels) and $E_q^*$ (lower panels). Left panels: At $T=0.01$ for $g=0.05$ (solid bold line), $g=0.1, 0.15, 0.2, 0.25, 0.3, 0.35, 0.4, 0.45, 0.5, 0.6, 0.7, 0.79$  (solid line) and $g=0.8, 0.9$ (dashed line). Right panels: At $T=0.1$, for $g=0.25, 0.32, 0.33, 0.34, 0.35$ (solid line) and $g=0.36, 0.37, 0.38, 0.39, 0.4, 0.45$ (dashed line).}
    \label{fig2}
\end{center}
\end{figure}

The present study, concentrating on the strong coupling regime, deals with 
the nature of transition from the superfluid state of resonating fermion 
pairs to their phase uncorrelated Bose Metal state. The critical interaction strength $g_c$ as function of temperature $T$ is indicated by  
diamonds ($\diamondsuit$) in Fig. \ref{fig1}. Such a  Bose Metal consists  of current carrying states being  composed of bonding pairs without any contributions from single particle fermion states. The transition is characterized by  $\mu^*$ moving out of the fermionic band  $\e_{k}^*$. It is 
discontinuous for $T\leq T_0 \simeq 0.75$ (thick solid line in Fig. \ref{fig1}) and changes into a continuous transition  for $T>T_0$ (dashed line). The nature of the transition can be deduced from the renormalized fermionic dispersion  $\e_{k}^*$ which is shown in the upper panels of Fig. \ref{fig2} for various exchange interactions $g$ and temperatures $T=0.01$ and $T=0.1$. 

For $T=0.01$ and $g<g_c\simeq0.8$ (solid lines of the upper left panel
of Fig. \ref{fig2}), the fermionic excitations show a single band
separated by a gap which is reminiscent of a lower and upper
Bogoliubov band, when keeping only the parts with maximal spectral
weight. As $g$ increases, the gap gets bigger and the fraction of
coherent fermions which are separated from incoherent excitations by
the superconducting gap becomes smaller.  For $g>g_c$ (dashed lines),
the lower Bogoliubov band has disappeared and the upper Bogoliubov
band now plays the role of our effective fermionic band $\e_k^*$ which
has abruptly moved up in energy and now lies above $\mu^*$. The fermionic
spectral function now consists of purely incoherent contributions
characterized by breaking up the strongly bound fermion pairs in the
purely local-phase correlated liquid. The total number of coherent and
incoherent fermions is approximately constant since the exchange
coupling $g$ does not significantly change the relative number of
bosons and fermions. For $T=0.1$, the change of the fermionic
dispersion is continuous for increasing $g$ when $\mu^*$ comes to lie
below $\e_{k=0}^*$, as shown on the upper right hand side of
Fig. \ref{fig2}. This yields $g_c\simeq0.35$.

\begin{figure}[t]
  \begin{center}
    \includegraphics*[width=3in,angle=0]{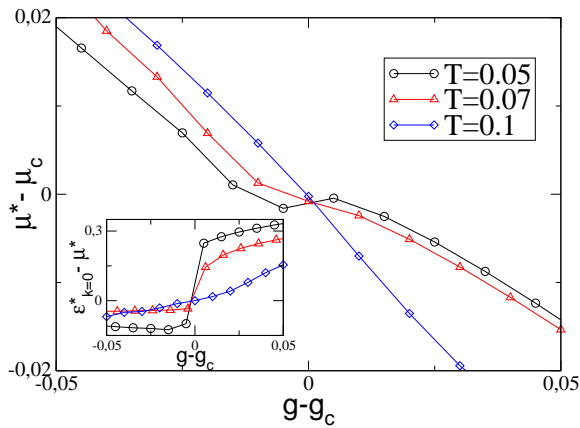}
    \caption{(Color online) The renormalized chemical potential $\mu^*$ as function of the coupling strength $g$ with respect to the critical values $\mu_c$ and $g_c$ of the phase transition from superfluidity to a Bose Metal. The curves correspond to three different temperatures $T>T_0$, $T\approx T_0$ and $T<T_0$ where $T_0$ defines the transition from second to first-order phase transition.}
    \label{fig3}
\end{center}
\end{figure}

At low temperatures and close to the phase transition, we notice a qualitative change in $\mu^*$ as a function of $g$ for different temperatures. Fig. \ref{fig3} shows $\mu^*$ as a monotonically decreasing function of $g$ for $T=0.1$, changing to a non-monotonic behavior at $T=0.5$. This is indicative for the continuous transition for $T>T_0\simeq0.75$ changing into a first order transition for $T\leq T_0$ where we find the coexistence of two phases with different relative densities of bosons and fermions. This feature is accompanied by a smooth, respectively abrupt, change of the position of $\mu^*$ from inside the  fermionic band $\e_k^*$ for $g \leq g_c$ to outside of that band for $g >g_c$, as seen in the insert of Fig. \ref{fig3}, where $\e_{k=0}^*-\mu^*$ is plotted as function of $g$.

For $T<T_0$, also the bosonic dispersion changes abruptly from a linear to a quadratic spectrum at low $q$ as $g$ increases beyond $g_c$ (lower left panel of Fig. \ref{fig2}). The initially dispersionless bosons thus first become superfluid and then itinerant as $g$ increases from weak to strong coupling.
For $T_0<T=0.1$ (lower right panel), the transition is continuous, nevertheless the width  of the occupation number $n_q^{BE}(\ell=\infty)$ around $q=0$ of the renormalized dispersion changes rapidly, indicating the transition from superfluidity to a Bose metal at finite temperatures.  

{\it Summary and discussion.} We have discussed the boson-fermion model within the flow equation approach, including both the Cooper-pair and the density-density channel for all wave numbers. We find a first order phase transition from a superfluid phase to a Bose Metal for low temperatures in the strong exchange coupling regime. In the fermionic one-particle spectral function, both phases show a gap. But, depending on the relative strength of the hopping matrix element $t$ with respect to the $g$, this leads either to a phase-correlated or to a phase-uncorrelated bonding-pair liquid. The transition is seen in the bosonic dispersion as a change from linear to quadratic in momentum behavior, i.e., in the superconducting regime we have the Goldstone mode $E_q^*-E_{q=0}^*=v_{\rm SF}q$ (for small $q$), whereas in the phase-uncorrelated regime we have $E_q^*-E_{q=0}^*=q^2/2m^*$ with $m^*\simeq0.1/t$. 

In the superfluid phase, the renormalized dispersion for the bosons is linear for wave numbers $q=0$ up to some  $q_{\rm max}$ which depends on the size of the regions where local phase correlations between the itinerant fermions and the tightly bound boson pairs prevail. They become stronger for $g\rightarrow g_c$ which renders impossible the maintenance of the spatial phase correlations for the Cooper-pairs in the fermionic subsystem over short distances and hence a linear dispersion of the bosons for large momenta. $q_{\rm max}$ thus decreases with increasing $g$ and eventually reaches zero at the phase transition. We note that the slope of the linear dispersion, i. e., the superfluid velocity $v_{\rm SF}$, hardly changes even for $g\simeq g_c$. This is yet a further manifestation of the first order phase transition (in case of a second order phase transition the slope would eventually tend to zero).

In the strong-coupling Bose metal phase $(g>g_c)$, all one-particle fermion states exclusively exist as components of the bonding pairs. The single-fermion particle spectrum then exhibits a correlation gap and the strong local phase coherence, leading to a locally fluctuating field,  destroys all coherence of the itinerant fermions. The bosonic dispersion, $E_q^*$, of the renormalized Hamiltonian corresponds, within such a flow equation procedure, to renormalized bosonic operators of the form $u_qb_q^\dagger+\sum_kv_{k,q}c_{k+q,\uparrow}^\dagger c_{-k,\downarrow}^\dagger$ \cite{DR0304}. Since  $E_q^*\propto q^2$, it represents a Bose Metal composed of the above mentioned bonding pairs. 

The first order transition has been deduced from the dynamical properties of the system. In order to discuss the transition within thermodynamic quantities, one has to study the flow of the operators in addition to the one of the Hamiltonian. This is a separate problem which will be addressed in the future. The present study dealt with spatially  homogeneous systems, but should have its bearing on inhomogeneous systems, such as fermionic gases in optical traps.

{\it Acknowledgements.} T.S. is grateful for the hospitality of the Institut NEEL, CNRS in Grenoble and support from MCyT (Spain) through grant FIS2004-06490-C03-00 and the Juan de la Cierva program. 


\begin{thebibliography}{99}
\bibitem{Comment} 
One can generalize Eq. (\ref{SCmu})  by multiplying the bosonic density 
$n_q^{BE}(\ell)$ by a parameter $\gamma(\ell)$, thus changing the relative 
weight with respect to the non-interacting case due to interaction effects.
Setting $\gamma(\ell)=1+\gamma_0e^{-D^2\ell}$ with $\gamma_0\in[-0.5,1]$, this leads to qualitativly the same phase diagram.
\end{thebibliography}
\end{document}